\documentclass[ twocolumn,
aps,nofootinbib,
showpacs,showkeys,preprint
tightenlines,preprintnumbers,] {revtex4}

\usepackage{epsf,epsfig,subfigure,
graphicx,amsmath,amssymb}
\usepackage{color}

\begin{document}
\newcommand{\dis}[1]{\begin{equation}\begin{split}#1\end{split}\end{equation}}
\def\ie{{\it i.e.~}}
\def\etal{{\it et al.\,}}
    
   \def\thb{\overline{\theta}}
  \def\Uanom{U(1)$_{\rm anom}$} 
       \def\NDW{N_{\rm DW}}
 \def\Z{{\bf Z}}
 \def\gev{\rm GeV}
\def\EE8{{\rm E}_8\times {\rm E}_8^\prime}
\def\Qanom{Q_{\rm anom}}

\title{'t Hooft mechanism, anomalous gauge U(1),  and ``invisible'' axion from string}\thanks{Talk presented at Patras 17, Thessaloniki, Greece, 15 May 2017.}

\author{{\slshape  Jihn E. Kim}\\[1ex]
Center for Axion and Precision Physics (IBS), 291 Daehakro, Daejeon 34141,  and\\
 Department of Physics, Kyung Hee University, 26 Gyungheedaero,  Seoul 02447,  Korea.}
 

\begin{abstract}
Among solutions of the strong CP problem, the ``invisible'' axion  in the narrow axion window is argued to be the remaining possibility among natural solutions on the smallness of $\bar{\theta}$. Related to the gravity spoil of global symmetries, some prospective ``invisible'' axions from theory point of view are discussed. In all these discussions including the observational possibility and cosmological constraints, including a safe domain wall problem, must be included. 
\end{abstract}

\maketitle

\section{The 't Hooft mechanism}\label{subsec:Hooft}

In the bosonic collective motion in the Universe \cite{KimYannShinji}, so far the ``invisible'' QCD axion \cite{KSVZ1} seems the mostly scrutinized one because it can also provide cold dark matter (CDM) in the Universe. It is a pseudo-Goldstone boson arising from spontaneous breaking of a global symmetry \cite{KimIJMP17}. In deriving a global symmetry in the process of spontaneous symmetry breaking, realizing the 't Hooft mechanism is crucial   \cite{Hooft71}, which is stated as, {\it ``If both a gauge symmetry and a global symmetry are broken  by the vacuum expectation value (VEV) of a complex scalar field, then the gauge symmetry is broken and one global symmetry remains unbroken.''}

Let us consider two phase field directions: one is the phase of a gauge U(1)$_1$ and the other  is the phase of a global U(1)$_2$ with two corresponding symmetry generators $Q_{\rm gauge}$ and $Q_{\rm global}$.  
 The proof of the 't Hooft mechanism is a very simple and elementary. It is obvious that the gauge symmetry is broken  because the corresponding gauge boson obtains mass.  Namely, only one phase or pseudoscalar is absorbed to the gauge boson, and there remains one continuous direction corresponding to the remaining continuous parameter.   To see this clearly, let us introduce a field $\phi$ on which charges $Q_{\rm gauge}$ and  $Q_{\rm global}$ act. The gauge transformation parameter is a local $\alpha(x)$ and the global transformation parameter is a constant $\beta$. Transformations are
\begin{eqnarray}
 \phi\to e^{i\alpha(x) Q_{\rm gauge}} e^{i\beta Q_{\rm global}}\phi,
\end{eqnarray}
which can be rewritten as
\begin{eqnarray}
 \phi\to e^{i(\alpha(x)+\beta) Q_{\rm gauge}} e^{i\beta (Q_{\rm global}-Q_{\rm gauge} )}\phi.
\end{eqnarray}
Redefining the local direction as $\alpha'(x)=\alpha(x)+\beta$, we obtain the transformation
\begin{eqnarray}
 \phi\to e^{i \alpha'(x) Q_{\rm gauge}} e^{i\beta (Q_{\rm global}-Q_{\rm gauge} )}\phi.
\end{eqnarray}
So, the $\alpha'(x)$ direction becomes the longitudinal mode of heavy gauge boson. 
Now, the charge $Q_{\rm global}-Q_{\rm gauge}$ is reinterpreted as the new global charge and is not broken by the VEV, $\langle\phi\rangle$, because out of two continuous directions one should remain unbroken. Basically the direction $\beta$ remains as the unbroken continuous direction. 
 
\section{The domain wall problem in ``invisible'' axion models}\label{sec:DWs}

It is well-known that if a discrete symmetry is spontaneously broken then there results   domain walls in the course of the Universe evolution. For the ``invisible'' axion models, it was pointed out that the domain wall number $\NDW$ different from 1 must have led to  serious cosmological problems in the standard Big Bang cosmology \cite{Sikivie82}. Therefore, the standard DFSZ models with $\NDW=6$ has not worked successfully in our Universe. We consider only $\NDW=1$ models  for  ``invisible'' axions.  
\begin{figure}[!h]
\centerline{\includegraphics[width=8cm]{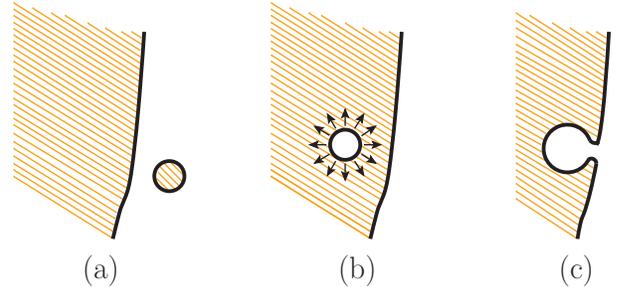} }
\caption{A horizon scale string-wall  for  $\NDW=1$ with a small membrane bounded by string.}\label{Fig:DWon}
\end{figure}
\begin{figure}[!t]
\centerline{\includegraphics[width=5cm]{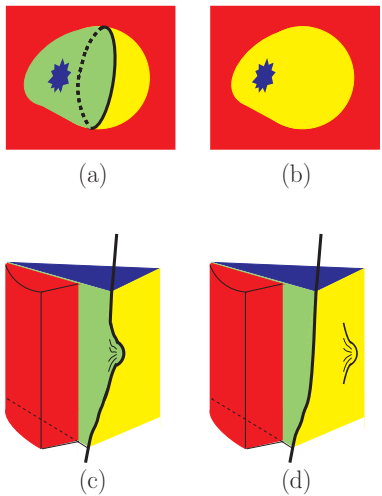} }
\caption{ For $\NDW=2$, small DW balls  (a) and (b), with punches showing the inside blue-vacuum, and the horizon scale string-wall system (c) and (d). Yellow walls are $\theta=0$ walls, and  yellow-green walls are $\theta=\pi$ walls.  Yellow-green walls of type (b) are also present.}\label{Fig:DWtw}
\end{figure}
The argument goes like this. In the evolving Universe, there always exists a (or a few) horizon scale string(s) and a giant wall attached  to it as shown in
Fig. \ref{Fig:DWon}\,(a). There are a huge number of small walls bounded by an axionic string which punch holes in the giant walls as shown in Fig. \ref{Fig:DWon}\,(b). The punched holes expand with light velocity and eat up the  giant string-walls as shown in
Fig. \ref{Fig:DWon}\,(c). This is the scenario that ``invisible'' axion models with $\NDW=1$ are harmless in cosmolgy. However, ``invisible'' axion models with $\NDW\ge 2$ have cosmological problems. 
For example for $\NDW=2$, a horizon scale string and wall system has the configurations shown in Figs. \ref{Fig:DWtw}\,(a), (b), (c), and (d), and one can see that  the horizon scale string-wall system is not erased.  
 With inflation, the domain wall problem has to be reconsidered as discussed in \cite{KimIJMP17}.  
  
\begin{figure}[!t]
\begin{center}
\includegraphics[width=0.5\textwidth]{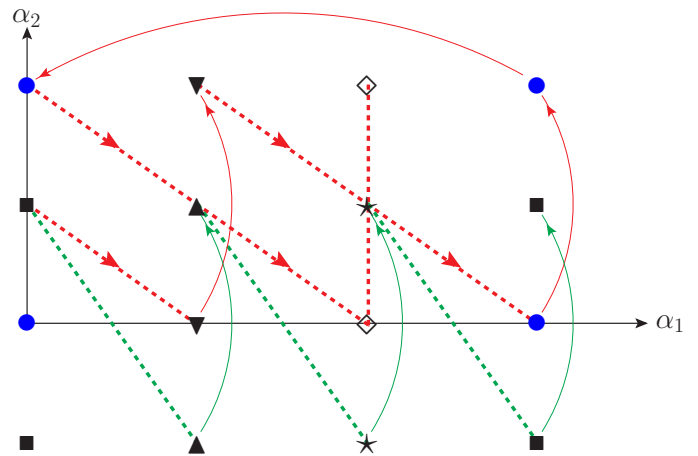} 
\end{center}
\caption{The Goldstone boson direction out of two pseudoscalars for $N_1=3$ and $N_2=2$ \cite{Choi85}.} \label{fig:GoldDirection}
\end{figure}

Thus, it is an important theoretical question, ``How can one obtain a reasonable ``invisible'' axion model having $\NDW=1$?''   One obvious model is the KSVZ axion with one heavy quark. Another more sophisticated solution is the Lazarides--Shafi (LS) mechanism in which the seemingly different vacua are identified by gauge transformation \cite{LS82}. The original suggestion was to  use the centers of extended-GUT groups for this purpose \cite{LS82}, which has been used in extended GUT models. But, the LS mechanism has a limited application.
More practical solutions come from using two discrete symmetry groups. This method can be extended to the Goldstone boson directions of spontaneously broken global symmetries \cite{Choi85}.  In Fig. \ref{fig:GoldDirection}, we consider two discrete symmetries with $\Z_3$ and $\Z_2$. $\alpha_1$ and $\alpha_2$ are the continuous directions. In case of Goldstone directions, two pseudoscalars are $a_1=f_1\alpha_1$ and $a_2=f_2\alpha_2$. If the potential is nonvanishing only in one direction,  then there is another orthogonal direction along which  the potential is flat. This flat direction is the Goldstone boson direction. For the discrete symmetry $\Z_3\times\Z_2$ of Fig. \ref{fig:GoldDirection}, there are six inequivalent vacua, marked as  blue bullet ($\color{blue}\bullet$), black down triangle ({\small$\blacktriangledown$}), diamond ($\diamond$), black square ({\tiny$\blacksquare$}), black triangle ({\small$\blacktriangle$}), and star ($\star$). Identifying along the red dash-arrow directions encompass all six vacua, where discrete group identifications along $\alpha_1$ and $\alpha_2$ are shown as horizontal and vertical arcs, respectively. Notice, however, if we identify along the green dash-arrow directions, then only {\tiny$\blacksquare$}, {\small$\blacktriangle$}, and $\star$ are identified. Parallel to the  green dash-arrow, there is another identification of $\color{blue}\bullet$,  {\small$\blacktriangle$}, and  $\diamond$. Thus, there remains $\Z_2$.
This example shows that all Goldstone boson directions are not necessarily identifying all vacua, even if $N_1$ and $N_2$ are relatively prime. The difference of red and green directions is a possibility of allowing a discrete group or not. In the red dash arrow case, when one unit of $\alpha_1$ is increased, one unit of $\alpha_2$ is increased. In this case, if  $N_1$ and $N_2$ are relatively prime, then all vacua are identified as the same vacuum, forbidding any remaining discrete group.  In the green dash arrow case, when one unit of $\alpha_1$ is increased, two units of $\alpha_2$ is increased, and $\Z_2$ is allowed in this direction. This fact was not noted in earlier papers  \cite{Choi85,KimPLB16}. Since the Goldstone boson is derived from VEVs of two Higgs fields in the above example, the possibility of identifying all vacua depends on the ratio of two VEVs. 

\section{Global U(1)  from anomalous gauge U(1) }\label{subsec:GlFromSt}

But, the most appealing solution is the direction related to the model-independent axion (MI-axion), $B_{\mu\nu}$,  arising from string compactification, which is known to have $\NDW=1$ \cite{Witten85,KimPLB16}. Even if one starts from string compactification, it is important to obtain $10^{9\,}\gev\le f_a\le 10^{11\,}\gev$. For this we need the 't Hooft mechanism with an anomalous gauge symmetry \Uanom. It is explicitly discussed in Ref. \cite{KimKyeNam17}.
It is generally known that 10 dimensional (10D) string theories do not allow global symmetries upon compactification, except the MI-axion direction,  $B_{\mu\nu}$.   Gauge fields arise from   compactification of $\EE8$ with gauge potential $A_M$. In the compactification of the heterotic string, the  MI-axion becomes the global shift direction \cite{WittenMI},
\dis{
a_{\rm MI}\to a_{\rm MI}+ {\rm constant}.\label{eq:MIshift}
}
where $a_{\rm MI}$ is the dual of the field strength of $B_{\mu\nu}$, $H_{\mu\nu\rho}=M_{\rm MI}
 \epsilon_{\mu\nu\rho\sigma}\,\partial^\sigma a_{\rm MI}$.
At the GUT scale, we need a global symmetry rather than the  MI-axion shift symmetry of (\ref{eq:MIshift}) such that the global symmetry is broken at the intermediate scale.  Here, the 't Hooft mechanism works.  
 $H_{\mu\nu\rho}$ couples to the non-Abelian gauge fields by the Green--Schwarz (GS) term \cite{GS84}. The GS term is composed of a product with one $B_{MN}$ and four non-Abelian gauge fields, $F_{PQ}, $ etc., contracted with $\epsilon^{MNPQ\cdots}$. 
 Thus, we obtain  
\dis{
\frac12 \partial^\mu  a_{\rm MI}\partial_\mu a_{\rm MI}+    M_{\rm MI}A_\mu\partial^\mu a_{\rm MI}+\frac12 M_{\rm MI}^2A^2_\mu.  \label{eq:HooftMI}
}
The GS term is generating the coupling $A_\mu\partial^\mu a_{MI}$. This coupling appears when there exists an anomalous \Uanom\, gauge symmetry from compactification of 10D $\EE8$ down to a 4D gauge group \cite{Anom87}. 

The  anomalous gauge symmetry \Uanom~is a subgroup of $\EE8$ with gauge field $A_\mu$.  One phase, \ie $\alpha=a_{\rm MI}/M_{\rm MI}$, is working for the 't Hooft mechanism, \ie the coupling $M_{\rm MI}A_\mu\partial^\mu a_{\rm MI}$ in Eq. (\ref{eq:HooftMI}), and hence one global symmetry survives below the compactification scale $M_{\rm MI}$.  The compactification scale $M_{\rm MI}$ is expected to be much larger than the decay constant $f_{\rm MI}$ of the MI-axion estimated in \cite{ChoiKimfa85}.
In the orbifold compactification, there appear many gauge U(1)'s which are anomaly free except the \Uanom. After the global \Uanom~is surviving below the scale $M_{\rm MI}$,   the 't Hooft mechanism can be applied repeatedly until all anomaly free gauge U(1)'s are removed around the GUT scale.  After making all these anomaly-free gauge bosons massive, the VEV $f_a$ of a SM singlet scalar $\phi$ breaks the global symmetry \Uanom~spontaneously and there results the needed ``invisible'' axion at the intermediate scale.  
 
Because it is so important to realize the intermediate scale $f_a$, let us discuss this string theory mechanism first in a hierarchical scheme and then present a general case based on the generalized 't Hooft mechanism.  In the literature, the Fayet--Iliopoulos terms (FI-term) for \Uanom~has been discussed extensively.  In the hierarchical scheme,  the VEVs of scalars is assumed to be much smaller than the string scale. Then, one can consider   the global symmetry \Uanom, surviving down from string compactification. Even if one adds the FI-term  for \Uanom, $|\phi^* \Qanom \phi-\xi|^2$ with $\xi\ll M_{\rm string}^2$, it is not much different from considering the global symmetry with the usual D-term,  $|\phi^* \Qanom\phi |^2$ (as if there is a gauge symmetry) since $\xi\ll M_{string}^2$. In fact, there exists a string loop calculation obtaining $\xi$ at string two-loop \cite{AtickTwoLoop88}. If the string two-loop calculation turns out to be hierarchically smaller (because of the two-loop) than the string scale, then the above  hierarchical explanation works. Even if the FI parameter $\xi$ is large, still there survives a global symmetry.   It is based on just counting the number of continuous degrees of freedom. Let us consider two phase fields, the MI-axion and some phase of a complex scalar $\phi$ carrying the U(1)$_{\rm anom}$ charge. Since we consider two phases and two terms, one may guess that the gauge boson ($A^\mu $) obtains mass and the remaining phase field also obtains mass by the FI D-term. But, it does not work that way, because there is no potential term for the phase fields to render such mass to the remaining phase field, because the charges of the gauge U(1) from $\EE8$ and the charge operator $\Qanom$ in the FI D-term are identical. It is equivalent to that there is no mass term generated because the exact Goldstone boson direction (the longitudinal mode of $A^\mu$) coincides with the phase of $\phi$ in the FI D-term.

 To discuss it explicitly, let us consider only one anomaly free U(1) gauge boson $A_\mu$ and the FI D-term for $\phi$ with generator $\Qanom$. Since $\phi$ carries the gauge charge $\Qanom$, we obtain its coupling to $A_\mu$ from the covariant derivative, by writing  $\phi=(\frac{v+\rho}{\sqrt2})e^{ia_{\phi}/v}$,
\dis{
 |D_\mu \phi|^2 &=|(\partial_\mu -ig
Q_{a}A_\mu)\phi|^2_{\rho=0}\\
&= \frac12(\partial_\mu a_{\phi})^2-gQ_a A_\mu \partial^\mu a_{\phi}+\frac{g^2}{2}Q_a^2v^2 A_\mu^2 .\label{eq:AmassPhi}
}
The gauge boson $A_\mu$ has the coupling to $a_{\rm MI}$ by the GS term, and the sum of two terms is
\dis{
 \frac12\left( g^2Q_a^2v^2 \right)&(A_\mu)^2+A_\mu( M_{\rm MI}\partial^\mu a_{\rm MI} -g Q_a v \partial^\mu  a_{\phi} )\\
 &+\frac12\left[ (\partial_\mu a_{\rm MI})^2+(\partial^\mu  a_{\phi})^2 \right].
\label{eq:scaleGl}   }
Thus, we note that $\cos\theta\,a_{MI} -\sin\theta\,a_{\phi} $ becomes the longitudinal degree of $A_\mu$ where
\dis{
 \sin\theta=\frac{g Q_a v}{\sqrt{M_{\rm MI}^2+g^2Q_a^2v^2}},
}
and a new global degree direction is
\dis{
\theta_{\rm QCD}\propto \cos\theta\,a_{\phi} +\sin\theta\,a_{\rm MI}.
 }
 $\theta_{\rm QCD}$ is the QCD vacuum angle direction and breaking \Uanom~at the intermediate scale produces the ``invisible'' axion. We obtained this important result from that only one combination of the phase fields is removed since the longitudinal degree of $A_\mu$ chooses the same generator for the shifts of $a_{\rm MI}$ and $a_\phi$.  Below the anomalous scale $\xi$, the 't Hooft mechanism is used repeatedly with anomaly free gauge U(1)'s together with the global \Uanom~we derived above, and there survives a global symmetry \Uanom~at the intermediate scale. Determination of  the VEV $f_a$ at the intermediate scale is such that the coefficient of $\phi^*\phi$ in the effective potential is given by $-(\rm intermediate~ scale)^2$. Note that choosing $\Qanom$ is not unique because one can add any combination of anomaly free gauge charges to $\Qanom$, without changing physics of \Uanom~\cite{KimPRD17}.

 \begin{figure}[!t]
\begin{center}
\includegraphics[width=0.4\textwidth]{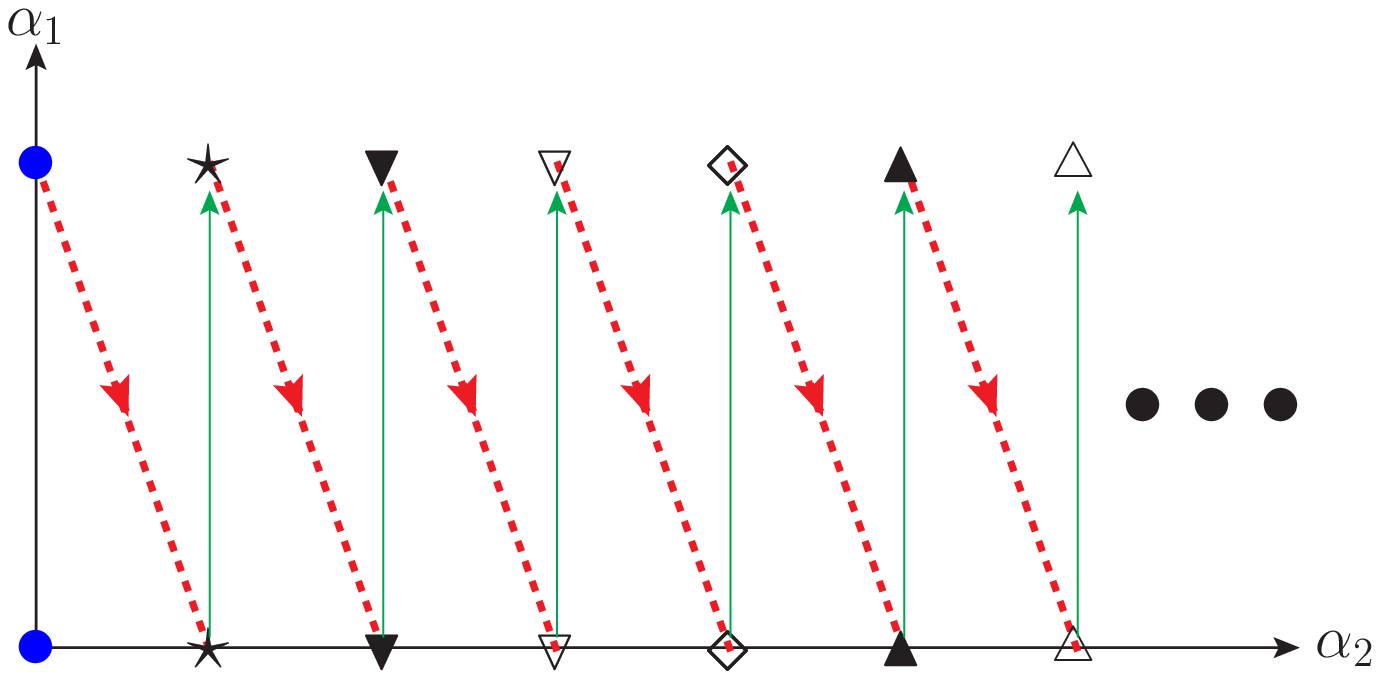} \\ (a)\vskip 1cm
\includegraphics[width=0.4\textwidth]{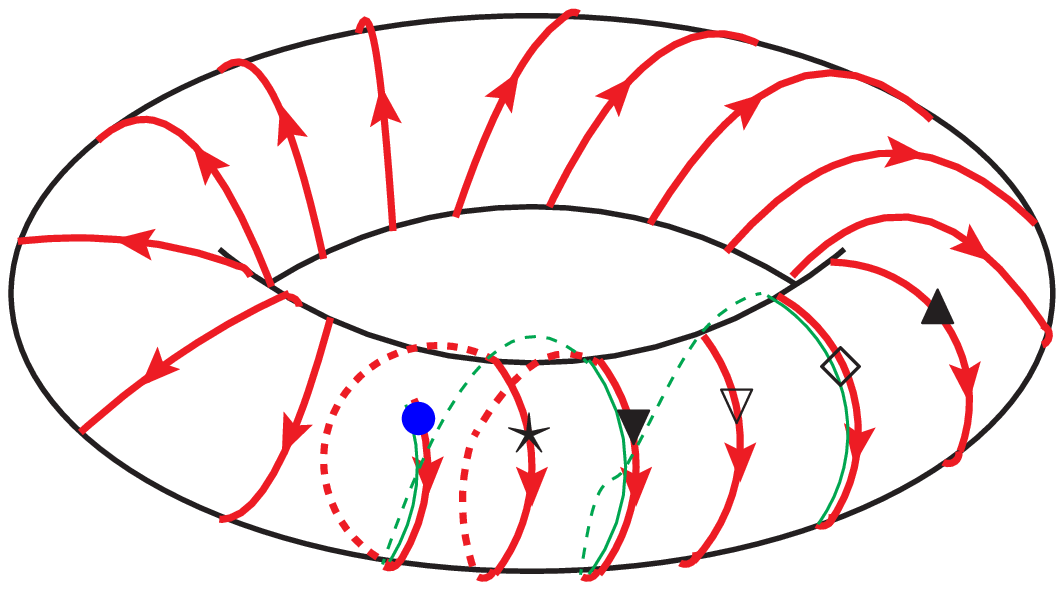} \\(b)
\end{center}
\caption{The MI-axion example of Fig. \ref{fig:GoldDirection}: (a) the standard torus identification, and (b) identification by the winding direction in the torus.} \label{fig:DWMIax}
\end{figure}

The global \Uanom~from string should not allow the cosmological DW problem of the ``invisible'' axion. The strategy with the Goldstone boson direction discussed in \cite{Choi85,KimPLB16} works here. For the Goldstone boson from the global \Uanom, we repeat it in
Fig. \ref{fig:DWMIax}\,(a). Since $\NDW=1$ in the MI-axion direction ($\alpha_1$  in Fig. \ref{fig:GoldDirection}), the red dash arrow direction identifies all vacua. In Fig.  \ref{fig:DWMIax}\,(b), it is re-drawn on the familiar torus. The red arrows show that $\alpha_2$ shifts  by one unit as $\alpha_1$ shifts one unit. In this case, all the vacua are identified and we obtain $\NDW=1$. The green lines show that $\alpha_2$ shifts  by two units for one unit shift of $\alpha_1$. If $N_2$ is even, then we obtain $\NDW=2$ since only halves of $N_2$ are identified by green lines. To find out $\NDW$, it is useful to factorize $N_2$ in terms of prime numbers.  Even though $N_2$ is very large, of order $10^3$, there are plenty of relatively prime numbers from those factors in $N_2$.  Figure  \ref{fig:DWMIax}\,(b) is drawn with $N_2=17$ and the green lines also identify all vacua since 1 and 17 are relatively prime. In string compactification, it is easy to find many relatively prime numbers  to all prime numbers appearing in the factors  of Tr\,$\Qanom$. For example, in Refs. \cite{KimPRD17,KimKyeNam17},   Tr\,$\Qanom$ was cited as --3492 which is expressed in terms of prime numbers as $2^2\times 3^2\times 97$. Not to introduce a fine-tuning on the ratio of VEVs, if we consider two VEVs are comparable, let us look for prime numbers relative to 2, 3, and 97. Near 3492, there are 3491, 3493, 3497, 3499, etc., relatively prime to 2, 3, and 97. So, a VEV of $\phi$ near the string scale, $\langle \phi\rangle=v_\phi/\sqrt2
$, and the longitudinal \Uanom~degree parameter $M_{\rm MI}$ render a global symmetry below the compactification scale such that the global current satisfies 
\dis{
\partial_\mu J^\mu_{\rm PQ}=\frac{1}{32\pi^2}
G^a_{\mu\nu} \tilde{G}^{a\,\mu\nu}.
}
Namely, two comparable VEVs,  $ v_\phi:M_{\rm MI} = 3491:3493$ for example, produce an exact global symmetry \Uanom~below the compactification scale.
If we neglect anomaly free gauge U(1)'s, a scalar field $\phi$ carrying the PQ charge houses the ``invisible axion'' with decay constant $f_a$ if $\langle\phi\rangle=f_a/\sqrt2$. $f_a$ is not the large ones such as $ v_\phi$ and $M_{\rm MI}$. So, the intermediate scale $f_a$ is not considered to be a fine tuning.  $f_a$ can come from another mechanism such as the supergravity scale \cite{KimScale84} or by some solution of the gauge hierarchy problem.

In conclusion, forbidding the DW problem for the ``invisible'' axion from the global \Uanom~is not considered as a fine-tuning  on the ratio of VEVs of Higgs fields. Due to the 't Hooft mechanism, the ``invisible'' axion scale can be lowered from the string scale down to an intermediate scale  \cite{KimKyeNam17}.
 
\section*{Acknowledgments}
This work is supported in part by the National Research Foundation (NRF) grant funded by the Korean Government (MEST) (NRF-2015R1D1A1A01058449) and by Institute of Basic Science of Korea (IBS-R017-D1).
  
\begin{footnotesize}

\end{footnotesize}
\end{document}